\documentclass[11pt,twoside]{article}

\usepackage{asp2006}
\usepackage{epsf}
\usepackage{lscape}

\begin{document}
\title{From deterministic to probabilistic population synthesis (why synthesis models are not what we thought they were, and how they can be much more than that)}   
\author{V. Luridiana and M. Cervi\~no}
\affil{Instituto de Astrof\'\i sica de Andaluc\'\i a (CSIC)}    

\begin{abstract} 
For a number of reasons, the properties of integrated stellar populations are distributed. Traditional synthesis models usually return the mean value of such distribution, and a perfect fitting to observational data is sought for to infer the age and metallicity of observed stellar populations. We show here that, while this is correct on average, it is not in individual cases because the mean may not be representative of actual values. We present a simple mathematical formalism to derive the shape of the population's luminosity distribution function (pLDF), and an abridged way to estimate it without computing it explicitly. This abridged treatment can be used to establish whether, for a specific case, the pLDF is Gaussian and the application of Gaussian tools, such as the $\chi^2$ test, is correct. More in general, our formalism permits to compute weights to be attributed to different properties (spectral features or band luminosities) in the fitting process. We emphasize that our formalism does not supersede the results of traditionaly synthesis models, but permits to reinterpret and extend them into more powerful tools. 
The reader is referred to the original paper for further details.
\end{abstract}

\section{Introduction }

When analyzing the integrated light of stellar populations, one usually wants to infer the population birth parameters (typically metallicity and age) from the observational data. Doing this is solving a statistical problem; the way it is generally done is by comparing the observational properties to predicted (that is, synthetic) properties. If the two coincide, it is assumed that the observed population's birth parameters are the same as those of the synthetic population. In the practice, this is done by placing the observational point on a grid of synthetic results.

This procedure is straightforward but also misleading, as it would be deducing the nationality of a woman by comparing the number of children she has to national natality statistics (which are an empirical rather than a synthetic grid, but this difference is irrelevant for the scope of our analogy). In this case, we readily recognize that the result would be, in general, incorrect, because of the large dispersion in the number of siblings around the mean national one: experience tells us that the dispersion around the mean value is of the same order of the mean value itself, so that the number of children is a really bad nationality indicator (with other indicators, such as the height or somatic traits, things are slightly better, so that we venture from time to time to make guesses on the base of them).

In the case of synthetic populations, we are not accustomed to apply the same line of reasoning. Although the luminosity predicted by a synthesis code is often interpreted as a deterministic quantity, it is, in fact, the average of the distribution of all the possible luminosities of clusters with given birth parameters. This basic fact is often overlooked and the `perfect fitting' is sought for. Yet this is dangerous if we don't know how large the variance of the distribution is, or whether the distribution itself is Gaussian - in most cases we don't even know whether it is symmetric! It is clear that, lacking information on the shape of the distribution, perfect fitting is not a necessary nor a sufficient condition for a correct inference. It is also clear that, if the property we are fitting is distributed, a range of possible solutions rather than a single one is all we can aspire to.

As for the reason that causes cluster luminosities to be distributed, there are several of them: the most obvious is the IMF sampling, but many others exist such as stellar rotation, differential mass loss, etc. Generally speaking, any fuzzy phenomenon at the stellar level will be reflected in a distribution at the cluster level. 

In this contribution, we want to help laying the statistical problem mentioned at the beginning on a proper basis. To do so, we have first to characterize this distribution, i.e. solve the probabilistic problem. In the remainder of this contribution, I will briefly mention previous attempts at solving the same problem by means of Monte Carlo simulations (Sect.~\ref{sec:MC}); introduce our probabilistic formalism and compare its performance to that of Monte Carlo simulations (Sect.~\ref{sec:probabilistic}); give two examples of applications (Sect.~\ref{sec:applications}); discuss the implications and draw our conclusions (Sect.~\ref{sec:conclusions}). A more detailed exposition can be found in \citet{CL06}. 
 
A few shortcuts and definitions will be used throughout the text. `Luminosity' indicates a luminosity in whatever wavelength or band (i.e. it can be monochromatic, in a band, or bolometric). The distribution of possible luminosity values of an individual star selected at random from the IMF is the stellar luminosity distribution function, or sLDF. The distribution of possible luminosity values of a stellar population is the population luminosity distribution function, or pLDF. `Cluster' and `stellar population' will be used interchangeably.

\section{Monte Carlo}\label{sec:MC} 
Previous attempts to characterize the pLDF have relied on the Monte Carlo method (e.g., \citet{B02}). Synthetic clusters are created through Monte Carlo sampling of the stellar masses entering the cluster. This approach has several disadvantages: i) it is time-consuming; ii) it is disk-space consuming iii) one must ensure that the result of simulations is stable; iv) the solution is not accompanied by any physical insight.

\section{The probabilistic formulation: a top-down approach}\label{sec:probabilistic}
It can be shown \citep{CL06} that the pLDF of
a cluster with given age and metallicity is given by the n-th convolution of the sLDF:

\begin{equation}
\varphi_{\mathrm{L_{tot}}}({\cal L}) = \overbrace{ \varphi_{\mathrm{L}}(\ell) \otimes \varphi_{\mathrm{L}}(\ell) \otimes\, ... \, \otimes \varphi_{\mathrm{L}}(\ell)}^{N_\mathrm{tot}}.
\end{equation}

The advantage of this formulation is readily seen: all the information is in fact enclosed in the sLDF, which we only have to convolve ${N_\mathrm{tot}}$ times. The huge sets of Monte Carlo simulations are substituted by a unique mathematical operation, the convolution (which is not, unfortunately, technically simple: see \citet{CL06}).
  
\subsection{Comparison with Monte Carlo results}   
The power of the convolution method is shown in Fig. 1. 
In the upper left panel, we show in grey shade the distribution of the $K$ luminosity of 1000 Monte Carlo simulations of 5.5 Ma clusters with 1 star, i.e. a Monte Carlo approximation to the sLDF. The different components of the histogram can be readily interpreted as the main sequence (the bunch at $0 < K \la 100$) and the post MS (the peak at high $K$), plus a 
contribution at $K=0$ from dead stars. 
In the same panel, we draw with a solid line an analytical approximation to the distribution made up of three Gaussians (the choice of adopting a Gaussian for each of the three components is motivated merely by computational easiness).

In the remaining left panels, independent Monte Carlo simulations for clusters with the same parameters and increasing ${N_\mathrm{tot}}$ are shown. Superimposed on each of the Monte Carlo histograms is a curve obtained by convolving the analytical curve of the first panel ${N_\mathrm{tot}}$ times with itself. The important thing to note here is that, thanks to the similarity between the analytical and the Monte Carlo sLDFs in the first panel, successive convolutions nicely reproduce the general patterns of the Monte Carlo simulations (recall the simulations are all independent!). Furthermore, we now have a clue regarding the multiple peaks seen in Monte Carlo simulations: these are not numerical instabilities, as the authors believed in a previous work, but rather the fingerprint of the clump of post-MS stars.

The peaks progressively smooth out in both the analytical curve and the Monte Carlo simulations as ${N_\mathrm{tot}}$ increases, and the overall shape of the curve becomes more and more symmetric. In the panels on the right, in which only the convolutions are shown, it can be seen how the curve progressively approaches a Gaussian shape. This does not depend on the inital curve being made up of Gaussians, but is the more general result stated by the central limit theorem: any distribution with finite moments will tend to a Gaussian if convolved with itself a sufficient number of times. In terms of our problem, this means that  the pLDF of a cluster is a Gaussian if the number of stars it contains is large enough.

\setcounter{figure}{0}

 \begin{figure}[!ht]
 \plottwo{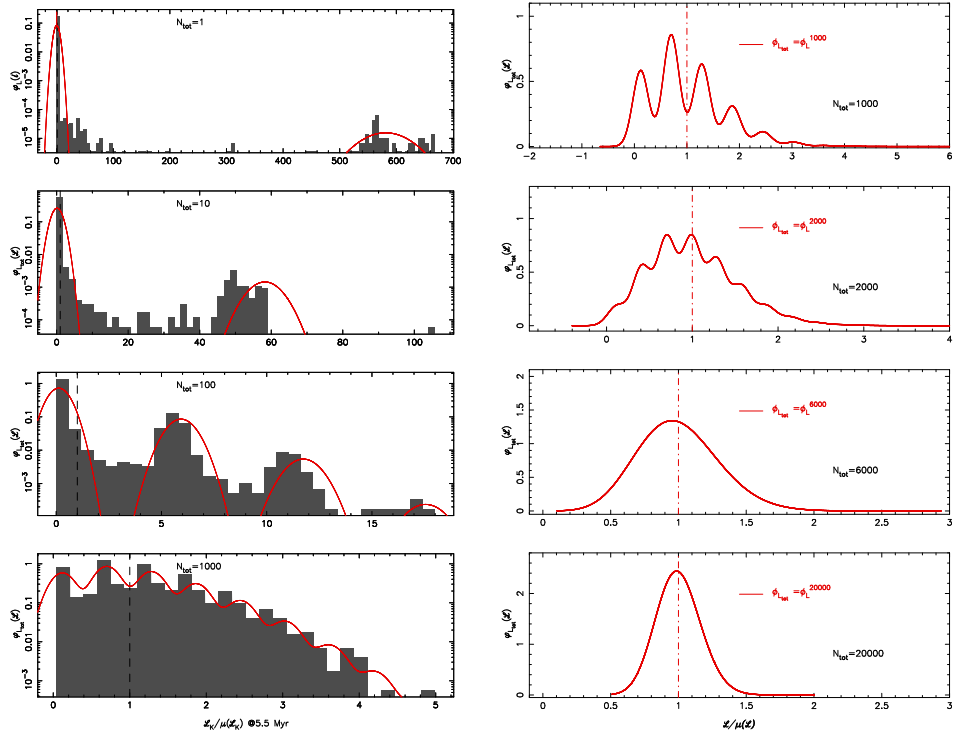}{luridiana_fig1b.eps}
 \caption{{\itshape Left:\/} $K$ luminosity of 5.5 Ma Monte Carlo clusters with different ${N_\mathrm{tot}}$ (grey shade histograms, compared with progressive convolutions of a simple sLDF (solid lines).
 {\itshape Right:\/} Convolutions as in the left panels, but with larger ${N_\mathrm{tot}}$ and plotted on a linear vertical scale.}
 \end{figure}

Before going on, let's recap here what we have found out until now: we have started by recognizing that the luminosity of a cluster with given age and $Z$ is distributed rather than univocally determined. We have seen that the distribution depends on the number of stars in the cluster ${N_\mathrm{tot}}$, and that it can be found by convolving the sLDF ${N_\mathrm{tot}}$ times with itself. Finally, we have seen that the pLDF of clusters with a large number of stars is Gaussian. How large is large enough? This will be the focus of next section.

\section{Applications of the probabilistic formulation: two examples}\label{sec:applications}  
As any distribution, the sLDF can be characterized by the mean and variance (the first and second raw moments), and the skewness and kurtosis (which are proportional to the third and fourth moments). 
Focusing on these quantities instead of the explicit expression is useful because simple scale relations, involving powers of ${N_\mathrm{tot}}$, hold between them and the corresponding quantities of a distribution obtained convolving the original distribution ${N_\mathrm{tot}}$ times. Therefore, the scale relations make it straightforward to derive the pLDF properties from the sLDF properties for any $N_{\mathrm{tot}}$ value.
Specifically,

\begin{eqnarray*}
\langle L \rangle & = &  N_{\mathrm{tot}}\times \langle l \rangle, \nonumber \\ 
\sigma^2(L) & = & {N_{\mathrm{tot}}}\, \sigma^2(l), \nonumber \\ 
\Gamma_1 & = & \frac{1}{\sqrt{N_{\mathrm{tot}}}}\, \gamma_1, \nonumber \\
\end{eqnarray*}
{and}
\begin{equation}
\Gamma_2  \, =  \, \frac{1}{N_{\mathrm{tot}}} \,\gamma_2.  
\end{equation}

\noindent where $\langle l \rangle$, $\sigma^2(l)$, $\gamma_1$, and $\gamma_2$ are the mean, variance, skewness, and kurtosis of the sLDF, and $\langle L \rangle$, $\sigma^2(L)$, $\Gamma_1$, and $\Gamma_2$ are those of the pLDF.
The first of the relations above gives mathematical foundation to the intuitive notion that the average luminosity scales linearly with ${N_{\mathrm{tot}}}$; this rule is routinely used when the output of traditional synthesis codes, which is the mean of the sLDF, is rescaled to the desired cluster size.
The second relation tells us that the relative dispersion decreases with increasing ${N_{\mathrm{tot}}}$. Finally, in the third and fourth relations we recover the central limit theorem: if $N_{\mathrm{tot}}$ is large, the skewness and kurtosis tend to 0, which are characteristic of a Gaussian. This is also a sufficient condition, and we can translate it into rough quantitative terms with the following prescription: a distribution is quasi-Gaussian within 3$\sigma$ from the mean if $\Gamma_1$, $\Gamma_2 < 0.10$.

\section{Applications of the probabilistic formulation: two examples}  
In addition to their standard output, the mean predicted luminosity, traditional synthesis codes can be easily adapted to yield the value of the remaining three quantities, that is the variance, skewness and kurtosis. In Fig. 2 (left) we see these four quantities of the sLDF in various bands as a function of age. 
Recalling the scale relations, these plots can be translated into their equivalent for clusters with any number of stars. This information, combined with the quantitative constraints mentioned at the end of the previous section, imply a lower limit on $N_{\mathrm{tot}}$ for Gaussianity. In the example above, the stronger constraint comes from the skewness plot, which implies that in the IR more than $10^6$ stars are needed for Gaussianity. 

A further example is given by the fitting of spectral lines. In the rightmost panels of Fig. 2 we see the same kind of plots as in the left panels, only that we have now represented the spectral distribution of a single model rather than a continuum of models at increasing ages. As examples of the use of these plots, it can be seen that the relative scatter is different in the K calcium lines, implying that they are not equally robust age indicators. The skewness and the kurtosis are higher in $H\delta$ than in the H and K calcium lines, meaning that the pLDF of this line is more asymmetric and wider than in the other lines. 

 \begin{figure}[!ht]
 \plottwo{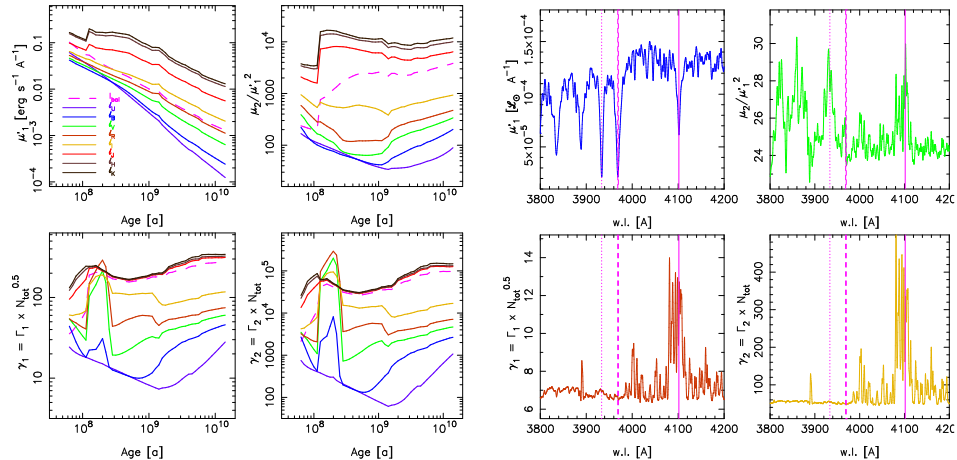}{luridiana_fig2b.eps}
 \caption{{\itshape Left:\/} Mean, variance, skewness, and kurtosis of the sLDF in different bands as a function of age.
 {\itshape Right:\/} Same as left, but for a single spectral energy distribution.}
 \end{figure}

\section{Conclusions}\label{sec:conclusions}  
We have shown here that, prior to solving the statistical problem of deducing population properties from the comparison with synthetic properties, one must solve the probabilistic problem of characterizing the distribution of properties as a function of birth parameters. The most straightforward and illuminating way of doing this is by computing the pLDF as the n-th convolution of the sLDF. Only after this step is achieved will we be able to make the reverse step of translating the observed properties in terms of inferred parameters. We can expect that no unique solution will be found, but rather a range of possible values and associated confidence levels. This is unfortunate but also inevitable, and we're better off if we recognize the inherent degeneracy of observed data rather than cling to an illusory determinism. As is often the case, the range of possible solutions will be tightened by means of the simultaneous use of several observables.

In the two practical examples seen above, it is shown that the mean, variance, skewness and kurtosis of the sLDF provide the necessary and sufficient information to do luminosity or spectral fitting of observed populations in a quantitative way, one in which each band or spectral feature is given a weight adequate to its robustness.

As a final remark, let us stress that our approach does not contradict the results of traditional synthesis models, but rather gives them a more precise physical interpretation and lays the basis to take full advantage of them; for example, while traditional synthesis models are not capable of tackling fuzzy phenomena (e.g., stellar rotation), these can be easily included in our formalism. More details can be found in \citet{CL06}.

\acknowledgements 
This work was supported by the Spanish {\it Programa Nacional de Astronom\'\i a y Astrof\'\i sica} through the project AYA2004-02703. MC is supported by a {\it Ram\'on y Cajal} fellowship. VL is supported by a {\it CSIC-I3P} fellowship.
A special thank is due to Cesare Chiosi for his pioneering contribution to this field and for a lovely conference.

\end{document}